\documentclass[sigconf]{acmart}

\AtBeginDocument{%
  }

\setcopyright{acmlicensed}
\copyrightyear{2018}
\acmYear{2018}
\acmDOI{XXXXXXX.XXXXXXX}

\acmConference[Conference acronym 'XX]{Make sure to enter the correct
  conference title from your rights confirmation emai}{June 03--05,
  2018}{Woodstock, NY}
\acmISBN{978-1-4503-XXXX-X/18/06}
\usepackage{multicol, multirow}
\usepackage{stfloats}
\usepackage{graphicx}
\usepackage{booktabs}
\usepackage{pgfplots}

\begin{document}

\title{Effective Knowledge Transfer for Multi-Task\\ Recommendation Models}

\author{Guohao Cai}
\authornote{Both authors contributed equally to this research.}
\affiliation{
  \institution{Huawei Technologies Co., Ltd.}
       \city{Shenzhen} 
   \country{China}
}
\email{caiguohao1@huawei.com}

\author{Jun Yuan}
\authornotemark[1]
\affiliation{%
  \institution{Huawei Technologies Co., Ltd.}
     \city{Shenzhen} 
   \country{China}
}
\email{yuanjun25@huawei.com}

\author{Zhenhua Dong}
\affiliation{%
  \institution{Huawei Technologies Co., Ltd.}
     \city{Shenzhen} 
   \country{China}
}
\email{dongzhenhua@huawei.com}



\begin{abstract}
The conversion rate (CVR) is a crucial metric for evaluating the effectiveness of platforms, as it quantifies the alignment of content with audience preferences. However, the limited nature of customers' conversion actions presents a significant challenge for training ranking models effectively. In this paper, we propose an Effective Knowledge Transfer method for Multi-task Recommendation Models (EKTM). This method enables the ranking model to learn from diverse user behaviors, thereby enhancing performance through the transfer of knowledge across distinct yet related tasks. Each specific CVR task can directly benefit from the insights provided by other tasks. To achieve this, we first introduce a router module that integrates and disseminates knowledge across tasks. Subsequently, each CVR task is equipped with a transmitter module that facilitates the transformation of knowledge from the router. Additionally, we propose an enhanced module to ensure that the transferred knowledge benefit the original task learning. Extensive experiments on several benchmark datasets demonstrate that our proposed method outperforms existing state-of-the-art approaches. Online A/B testing on a commercial platform has validated the effectiveness of the EKTM algorithm in large-scale industrial settings, resulting in a 3.93\% uplift in effective Cost Per Mille (eCPM). The algorithm has since been fully deployed across two of the platform’s main-traffic scenarios.
\end{abstract}


\maketitle


\section{Introduction}
\begin{figure}[!tbp]
    \centering
     \includegraphics[scale=0.23]{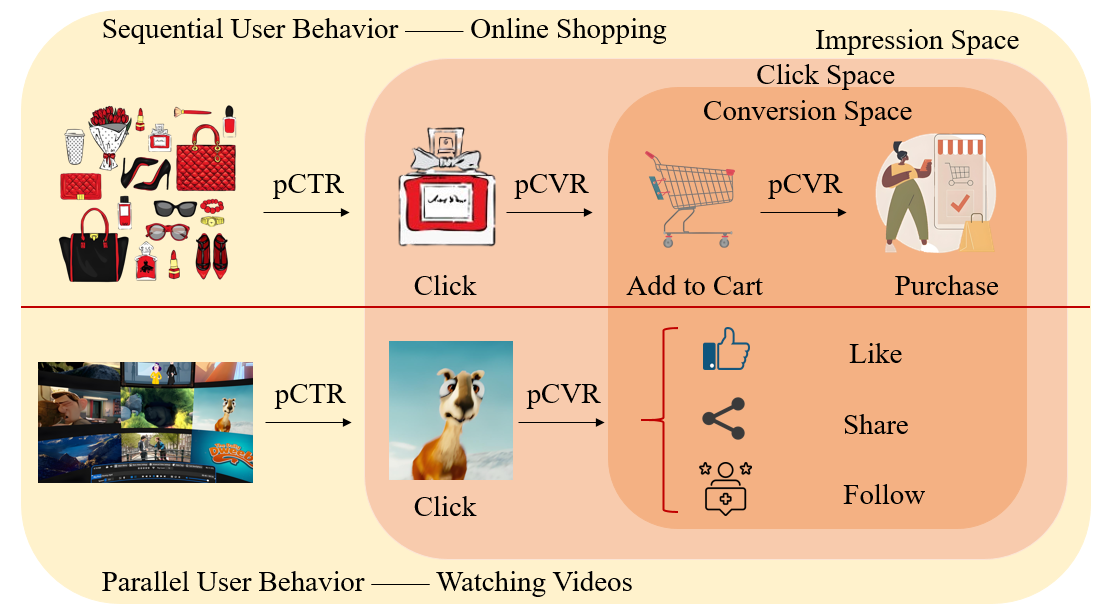}
    \caption{An illustrated example of two typical user behavior patterns in multi-task learning recommendation system.
    }
    \label{fig:intro}
\end{figure}

In the information-rich age, recommender systems, which deliver personalized content to users, are widely adopted and require considerable amount of effort to optimize on online platforms.  Typically, users conversion behaviors in multi-task learning (MTL) recommendation system could be categorized into two primary types: sequential pattern and parallel pattern, as shown in Figure ~\ref{fig:intro}. Firstly, user clicks an item from many exposed candidates ranked by a click prediction rate (CTR) model. Then multiple conversion actions could be take place in sequence such as add to cart and purchase in online shopping event; or in parallel such as give a like and follow the content creator in watching video event.


Compared to the massive impression data, the quantity of positive samples collected from users' click behavior logs is relatively small. Such issue become serious when it comes to conversion space, poses challenge to CVR prediction task, for the relative sparsity of training samples. MTL models partially address this issue by leveraging samples from impression space. Usually, these models \cite{ma2018mmoe,tang2020ple,AAAI24_STEM,DBLP:journals/ijon/WuFXYY25} adopt customized bottom and task-specific tower (each task maintains a separate tower that is optimized only with its corresponding loss) architecture. However these methods fails to take full advantage of knowledge from other tasks, since representations closer to the output layer contain richer information~\cite{DBLP:journals/corr/TzengHZSD14,DBLP:journals/corr/LiYCLH15}. On the other hand, knowledge transfer of scalar output ~\cite{ESMM,ESCM2} is limited when predicted indicators are not on the same scale, such as CTR prediction and watch time prediction. How to effectively integrate knowledge from other tasks and ensure positive transferring remains a significant challenge. 


In this paper, we introduce a novel method named as effective knowledge transfer method for multi-task recommendation models. It enables the ranking model to effectively learn from a diverse array of user behaviors, improving performance by leveraging the knowledge transferred from other tasks. To achieve this goal, we begin by designing a router module that serves to integrate and distribute knowledge among the different tasks. It is worth noticing that information from other tasks are not always bring advantage to current task learning, and may have adverse effects, namely, negative transferring~\cite{DBLP:series/sci/TorreySWM10,AAAI24_STEM}. We propose a transmitter module for each CVR task, which plays a vital role in inspecting and importing the relevant knowledge from the router. It ensures that only the most pertinent and valuable information is utilized. Furthermore, we propose an enhancing CVR prediction module that guarantee the transferred knowledge will contribute positively to the learning process of each task. 

The main contributions of this work are as follows:
\begin{itemize}
    \item To mitigate the data sparsity issue, we propose a centralized router module and a transmitter module that integrate and shares knowledge among different tasks. 
    \item To guarantee the positive direction of knowledge transfer further, an enhanced CVR prediction module is introduced to align with the specific learning objective of each task, ensuring that insights from related tasks boost individual task performance. 
    \item We carried out comprehensive experiments that involved the use of both public  datasets as well as an industrial platform. The results clearly illustrate and confirm the effectiveness of our approach in real-world scenarios and varied environments.
\end{itemize}


\section{Related works}

Multi-task learning has gained popularity in recommendation systems by enabling models to learn from multiple related tasks at once, enhancing generalization and performance. By leveraging knowledge transfer across tasks—both implicitly and explicitly—we present the key multi-task learning studies pertinent to our work.

\textbf{Knowledge Transfer Implicitly.}
From single task to multi-task learning, the intuitive method is to construct several towers for different tasks, and share the same bottom layer for feature interaction, reffed to Share Bottom method~\cite{Caruana97}. MFH~\cite{DBLP:conf/cikm/LiuLAXW22} leverages nested hierarchical MTL trees to capture multi-dimensional task relationships at a macro level. Recently, the expert-gate pattern has become the mainstream of the MTL in industrial field. The MMOE~\cite{ma2018mmoe} employs softmax-based gates within gating networks to assign varying weights to different experts, optimizing their combination for each specific task. PLE~\cite{tang2020ple} introduces a Customized Gate Control (CGC) module that differentiates between shared and task-specific experts. MoSE~\cite{DBLP:conf/kdd/QinCZCMQ20} captures sequential user behaviors by employing sequential experts and utilizing Long Short-Term Memory (LSTM) in both the shared bottom layer and the task-specific towers. Recently, STEM~\cite{AAAI24_STEM} propose a share and task-specific embedding approach to address the issue of negative transfer in MTL recommendations. However, these methods transfer shallow knowledge through shared parameters in the bottom layers, limiting the level of assistance across tasks. Besides, it is common for some specific tasks to be learned well while others are overlooked. 


\textbf{Knowledge Transfer Explicitly.}
Another form of knowledge transfer involves designing interactions among tasks explicitly. In ESMM~\cite{ESMM}, the output layers calculate the post-impression click-through and conversion rate by combining the probabilities of post-impression click-through rate and post-click conversion rate through multiplication. In ESM2~\cite{ESM2}, a user sequential behavior graph is constructed, with additional tasks decomposed to enable probability transfer. ESCM2~\cite{ESCM2} employs a counterfactual risk miminizer as a regularizer in ESMM to alleviate the estimation bias. These methods transfers only basic probability via the scalar product, ignoring richer representations in the vector space, resulting in significant losses. The most related work is AITM~\cite{AITM}, which model sequential dependencies in audience multi-step conversions. It assumes that the former task has a higher conversion probability than the latter, but this assumption may not hold true in all scenarios. For example, interest in an item may result in a high CVR, but if the item is not prominently displayed, the CTR could remain low.



\textbf{Transfer Learning}
Transfer learning in recommendation systems involves leveraging knowledge from one domain to improve performance in another domain. Public research works can be broadly categorized into three groups. Model-Based Transfer Learning~\cite{DBLP:conf/icml/LiYX09,DBLP:journals/corr/abs-2201-09679,DBLP:conf/icml/WalkerVLDAWH23}: transfer knowledge between different recommendation models, such as collaborative filtering, content-based filtering, and hybrid models. These approaches aim to transfer learned representations or parameters from one model trained on a source domain to a similar model for the target domain. Feature-Based Transfer Learning: transfer features extracted from one domain to another. For example, using user demographics or behavioral features learned from one domain to enhance recommendations in another domain with sparse data. Domain Adaptation Techniques: These methods focus on adapting the recommendation model from a source domain to a target domain while accounting for domain shift or differences in data distributions.

\begin{figure*}[t]
    \centering
    \includegraphics[width=0.95\textwidth,height=0.48\textwidth]{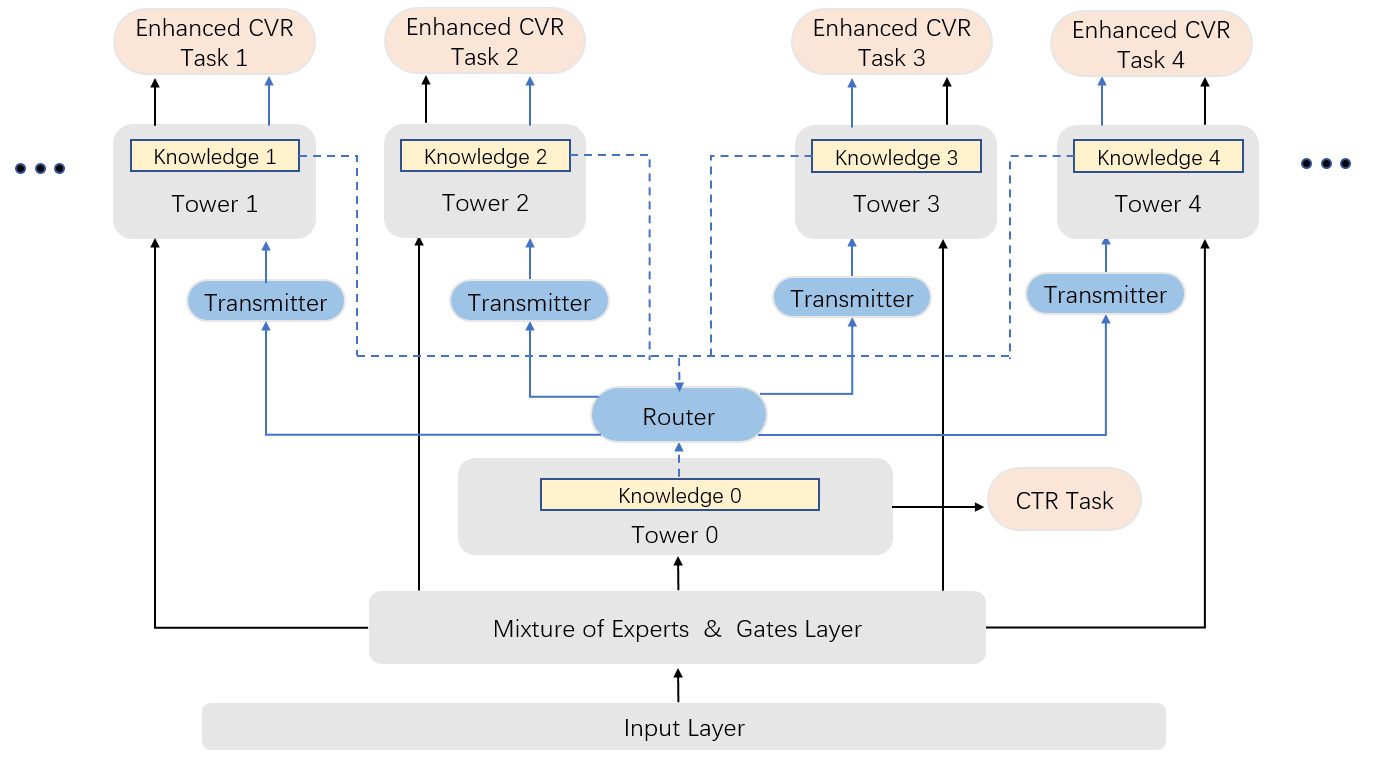}
    \caption{Overview of the EKTM architecture: The black arrows represent the data flow of the dominant architecture used in current industrial MTL models. The blue arrows highlight the innovative aspect of our approach, with dashed lines indicating the application of stop gradient.}\label{overview}
\end{figure*}

\section{Proposed Architecture}
Important concepts and notations are introduced  in Table~\ref{tab:notations} for clearly understanding. 

\begin{table}[htp]
\caption{Summary of notations}
\label{tab:notations}
\centering
\begin{tabular}{c|c}
\toprule
\textbf{Notation}&\textbf{Description}\\\hline
$T$&Number of the CVR tasks\\
$d$& Dimension of knowledge vectors\\
$m$& Number of selected tasks\\
$F_i$& Fusion knowledge to the $i$-th CVR task\\
$\mathcal{L}_{ctr}$ & Loss of CTR task\\
$\mathcal{L}^{i}_{cvr}$ & Loss of \textit{i}-th CVR task\\
$f_{ctr}$&The representation vector of CTR task\\
$f^{i}_{cvr}$&The vector for the \emph{i}-th CVR task.\\
\bottomrule
\end{tabular}
\end{table}
This section provides an in-depth description of EKTM method, which consists of three primary modules. As shown in Figure ~\ref{overview}, the model architecture consists of a CTR prediction task and several CVR prediction tasks. Firstly, input features are mapped into embedding vectors, which are then passed through mixture of experts equipped with gating mechanisms layer to facilitate feature interactions. Each task is assigned a dedicated tower with its own prediction layer. The data flow, represented by the black arrows, illustrates the conventional architecture of modern MTL models; while the data flow represented by the blue arrows are the designed components of our method. The dashed lines represent stop gradients, indicating where the gradients are halted during backpropagation. 


\subsection{Knowledge Transferring}

We first design a router module to achieve task-specific distribution, which routes the knowledge (hidden representation) from other tasks to the current one. Let $ f_{ctr} $ denote the knowledge vector from the CTR task, and $ f_{cvr}^i $ represent knowledge vector for the $i$-th CVR task. Take user behaviors in video surfing as example, knowledge from the 'like' task can benefit the learning process of 'share' task and 'follow' task. The correlations between knowledge of different tasks are calculated via the normalized cosine similarity, denoted as:
\begin{equation}
 s_{i,j} = \frac{f_{cvr}^i \cdot  f_{cvr}^j}{||f_{cvr}^i||_2 \times ||f_{cvr}^j||_2}, \quad
 \hat{s}_{i,j} = \frac{e^{s_{ij}}}{\sum_{j=1,j\neq i}^k e^{ s_{ij}}}
\end{equation}
To take advantage of the entire set of impression data and the correlations between CVR tasks, the router output for the $i$-th CVR task is organized as follow:
\begin{equation}
\begin{split}
    F_i = mean(f_{ctr}, \hat{s}_{i,1}f_{cvr}^1,..., \hat{s}_{i,i-1}f_{cvr}^{i-1},
    \hat{s}_{i,i+1}f_{cvr}^{i+1},...,\hat{s}_{i,T}f_{cvr}^T)
\end{split}
\end{equation}
where $T$ is the number of CVR tasks, and $d$ is the dimension of kowledge vectors.


Then we propose a transmitter module to incorporate complementary information from  that the current task needs. We utilize multi-head cross-attention to comprehensively select knowledge from the router. The attention mechanism uses queries and keys coming from $F_i$, while the values come from the current task $f_i$. To capture diverse patterns and relationships simultaneously, the multi-head cross-attention is adopted. Specifically, it is computed as follows: 
\begin{equation} MHCA(F_i)= \textit{Softmax}\left(\frac{(F_iW^Q)(F_iW^K)^T}{\sqrt{d}}\right)f_{cvr}^iW^V,
\end{equation} 
where $W^Q$, $W^K$, and $W^V$ are distinct learnable projection matrices and $d$ is the projected dimension. In the literature, several works have demonstrated that  the information from other tasks are not always bring advantage to current task learning. Inspired by the forget gate in Long Short-Term Memory (LSTM)~\cite{LSTM}, we equip the transmitter module with a gating mechanism to filter out the irrelevant information. 
\begin{equation}
   G_i =  \sigma( MHCA(F_i)^Tw_f  + b_f) \ast f_{ctr}
\end{equation}
where $w_f \in \mathbb{R}^{k\times 1}$ is the input weight and $b_f \in \mathbb{R}^{d\times 1}$ is the corresponding bias term. $\sigma$ is the sigmoid function and $\ast$ means the element-wise product. $G_i$ is the filtered knowledge for CVR task $i$, feed into the tower layers and get the regression loss or classification loss independently from original input.



%
\subsection{Enhanced CVR Prediction}
To safeguard the efficacy and positivity of knowledge transfer, A enhanced CVR prediction module is introduced. As shown in Figure~\ref{overview}, two types of arrows emanate from the tower layer to the prediction layer. The black arrows represent the calculation to the original losses, while the blue arrows correspond to the losses derived from the transferred knowledge. For each CVR task, we define the original loss as $L_o$ and the loss due to the transferred knowledge as $L_t$. The overall calibration loss for all tasks is then formulated as follows:
\begin{equation}
    L = L_{ctr} + \sum_{i=1}^{T}{((1-\alpha) L_o^{(i)} + \alpha max(0, L_t^{(i)} - L_o^{(i)}))}
\end{equation}
where $k$ is the total number of CVR tasks, hyper-parameter $\alpha \in (0,1)$ is for the balance of calibration loss and original loss. 

For the CTR task, a common objective function is the binary cross-entropy loss:
\begin{equation}\label{eq:ctr}
    L_{ctr} = \sum_i^n -y^ilog(p_{ctr}^i) - (1-y^i)log(1-p_{ctr}^i)
\end{equation}
where $n$ is the number of training samples, $p_{ctr}$ is the predicted click-through rate output by tower of CTR task. 

For the parallel CVR tasks, the collected signal could be continuous value metrics such as watch time, then the transferred knowledge loss $L_t$ and the original loss $L_o$ are typically defined using mean squared error or mean absolute error.
\begin{equation}
    L = \sum_i^n||p_{cvr}^i - y^i||^2
\end{equation}
Besides, the loss function is the same as Equation~\ref{eq:ctr} if the conversion binary signal is based on actions like "like" or "dislike".

For the sequential CVR tasks, we adopt the post-view click and conversion rate loss. This loss function leverages all samples from the impression space to instantiate both $L_o$ and $L_t$:
\begin{equation}
    L = -yzlog(p_{ctcvr}) - (1-yz)log(1-p_{ctcvr})
\end{equation}
where $y$ is the label denotes where the item is clicked, and $z$ is the label denotes where the item is converted. $p_{ctcvr}$ is the product of CTR prediction value and CVR prediction value.

\section{Experiments}

\subsection{Experimental Setup}

\subsubsection{Datasets.}\label{datasets}
We conduct experiments on two public datasets, and one industrial product dataset. Public dataset one AliExpress \footnote{https://tianchi.aliyun.com/dataset/74690} consists of user logs from real-world traffic on the e-commerce platform. We use data from four countries: US, ES, FR, and NL. The dataset includes two binary classification tasks: click-through rate prediction and post-click conversion rate prediction. The second public dataset KuaiVideo\footnote{https://www.kuaishou.com/activity/uimc/datadesc} has three tasks from three binary user behaviors: click, like, and follow. It is collected from a popular short video platform. The industrial dataset contains eight tasks, which is collected and sampled from company's product with more than 67M samples. The detailed information of public datasets is summarized in Table \ref{tab:aliexp} and Table \ref{tab:KuaiVideo}.

\begin{table}[h]
\caption{Statistics of the AliExpress dataset.}
\label{tab:aliexp}
  \centering
\begin{tabular}{cccccc}
\toprule
\toprule
& Split& Samples & CTR & CVR & CTCVR \\
\midrule
\multirow{2}{*}{US} & Train & 20M & 1.5\% & 2.41\% & 0.035\% \\
&Test& 7.5M & 2.2\% &2.41\% &0.052\% \\
\midrule
\multirow{2}{*}{ES} & Train & 22M & 2.6\% & 2.25\% & 0.58\% \\
&Test& 9.3M & 2.8\% &2.3\% &0.066\% \\
\midrule
\multirow{2}{*}{FR} & Train & 18M & 1.9\% & 2.63\% & 0.049\% \\
&Test& 8.8M & 2.2\% &2.71\% &0.061\% \\
\midrule
\multirow{2}{*}{NL} & Train & 12.2M & 2.0\% & 3.63\% & 0.073\% \\
&Test& 5.6M & 2.4\% &3.61\% &0.088\% \\
\bottomrule
\bottomrule
\end{tabular}
\end{table}

\begin{table}[h]
\caption{Statistics of the KuaiVideo dataset.}
\label{tab:KuaiVideo}
  \centering
\begin{tabular}{ccccc}
\toprule
Split& Samples & Click & Like & Follow \\
\midrule
Train & 10.9M &0.2058&0.0029& 0.0010  \\
Test  & 2.7M &0.1935 &0.0027&0.0010\\
\bottomrule
\bottomrule
\end{tabular}
\end{table}

\begin{table*}[tp]
\caption{Experimental on two-tasks Aliexpress dataset in four scenarios. The metric is AUC and reported Values are averages over 5 random seeds. The $^\ddag$ indicates that result is significantly outperform backbone with $p<0.05$ on t-test. The best results are shown in bold.}
\label{tab:two_tasks}
\centering
\begin{tabular}{l|cc|cc|cc|cc}
\toprule
\textbf{Scenarios}& \multicolumn{2}{c|}{US} &\multicolumn{2}{c|}{ES} & \multicolumn{2}{c|}{FR} & \multicolumn{2}{c}{NL} \\
\textbf{Tasks} & CTR & CTCVR & CTR & CTCVR & CTR & CTCVR & CTR & CTCVR \\\hline
\textbf{Single task Learning} &0.7058 &0.8637 &0.7252 & 0.8832 & 0.7174 &0.8702 & 0.7203 & 0.8556\\\hline
\textbf{Transfer Methods}& & & & & & &  \\
ESMM&0.7048&0.8710& 0.7273&0.8920& 0.7237&0.8714&0.7245&0.8613\\
AITM&0.7048&0.8730& 0.7290&0.8885& 0.7236&0.8763&0.7240&0.8577\\\hline
\textbf{MTL Backbones}& & & & & & &  \\
Shared-Bottom&0.7029&0.8698&0.7287&0.8866&0.7245&0.8700&0.7222&0.8590\\
PLE &0.7071&0.8732&0.7306&0.8930&0.7297&0.8780&0.7233&0.8639\\
STEM&0.7014&0.8704&0.7309&0.8922&0.7234&0.8751&0.7224&0.8537\\
MMOE&0.7092&0.8717&0.7290&0.8959&0.7274&0.8789&0.7247&0.8645\\\hline

\textbf{Ours}& & & & & & &\\
Shared-Bottom+EKTM&0.7053$^\ddag$&0.8711$^\ddag$&0.7299$^\ddag$&0.8885$^\ddag$&0.7251$^\ddag$&0.8715$^\ddag$&0.7244$^\ddag$&0.8609$^\ddag$  \\
PLE+EKTM &0.7096$^\ddag$&\textbf{0.8764$^\ddag$}&0.7318$^\ddag$&0.8959$^\ddag$ &\textbf{0.7315$^\ddag$}&0.8806$^\ddag$&0.7249$^\ddag$&0.8653$^\ddag$\\
STEM+EKTM  &0.7057$^\ddag$&0.8715$^\ddag$&\textbf{0.7321$^\ddag$} &0.8933$^\ddag$ &0.7261$^\ddag$ &0.8770$^\ddag$ &0.7233$^\ddag$&0.8607$^\ddag$ \\
MMOE+EKTM &\textbf{0.7139$^\ddag$}&0.8738$^\ddag$&0.7314$^\ddag$&\textbf{0.8971$^\ddag$}&0.7301$^\ddag$&\textbf{0.8810$^\ddag$}&\textbf{0.7263$^\ddag$}&\textbf{0.8694$^\ddag$}\\
\bottomrule
\end{tabular}
\end{table*}

\begin{table}[htp]
\caption{Ablation study on customs and calibration loss. AUC Values are averages over 5 random seeds. The best results in each column are bold with $p<0.05$.}
\label{tab:abl_tasks}
\resizebox{0.48\textwidth}{!}{
\begin{tabular}{l|ll|ll}
\toprule
\multirow{2}{*}{\textbf{Method}}& \multicolumn{2}{c|}{US} & \multicolumn{2}{c}{NL} \\
& CTR & CTCVR & CTR & CTCVR\\\hline
MMOE &0.7092&0.8717&0.7247&0.8645\\
\ with linear &0.7017 &0.8711&0.7200 & 0.8659  \\
\ with transmitter  &0.7102&0.8719&0.7255&0.8662\\
\ with linear+aux\_loss &0.7053 &0.8721 &0.7180 &0.8685   \\
MMOE+EKTM &\textbf{0.7139}&\textbf{0.8738}&\textbf{0.7263}&\textbf{0.8694}\\
\bottomrule
\end{tabular}
}
\end{table}

\subsubsection{Baselines.}\label{backbones}
We conduct experiments compare to six state-of-the-art multi-task learning methods in recommendation system:
\begin{itemize}
    \item ESMM is designed to accurately estimate post-click conversion rate by modeling user behavior sequentially across the entire impression space, i.e., from impression to click to conversion. It addresses two key challenges in CVR prediction: sample selection bias and data sparsity, by training directly on all impression samples instead of only clicked ones.
    \item AITM captures the sequential dependence between multiple conversion steps by adaptively transferring relevant information from earlier to later stages. It jointly models multiple conversion stages and addressing data sparsity and delayed feedback challenges common in targeted advertising scenarios.
    \item Shared-Bottom is one of the original neural network based multi-task learning methods, which has inspired researchers to devote efforts and proposed many wildly-used MTL methods.
    \item MMOE models task relationships in multi-task learning by employing multiple gating networks, each of which dynamically assembles shared expert networks to generate task-specific predictions. This structure enables the model to automatically learn both shared and task-specific patterns, improving performance especially when tasks have varying degrees of relatedness.
    \item PLE explicitly separates shared and task-specific components to avoid negative transfer and performance degradation. It employs a progressive routing mechanism with multi-level experts and gating networks to gradually extract and fuse deeper semantic knowledge, significantly improving joint representation learning and recommendation performance.
    \item STEM is a multi-task recommendation model that leverages shared and task-specific embeddings to effectively capture both common and unique patterns across tasks. By unifying representation learning with embedding interactions, STEM significantly enhances recommendation accuracy and efficiency in complex multi-task scenarios.
\end{itemize}

\begin{table*}[tp]
\caption{Experimental on three-tasks KuaiVideo dataset. The metrics are AUC and LogLoss, and the reported values are averages over 5 random seeds. The best results are shown in bold.}
\label{tab:three_tasks}
\centering
\begin{tabular}{l|cc|cc|cc}
\toprule
\textbf{Tasks}& \multicolumn{2}{c|}{Click} &\multicolumn{2}{c|}{Like} & \multicolumn{2}{c}{Follow}  \\
\textbf{Metircs}& AUC $\uparrow$ & LogLoss $\downarrow$ & AUC $\uparrow$ & LogLoss $\downarrow$ & AUC $\uparrow$ & LogLoss $\downarrow$ \\\hline
\textbf{Single task Learning} &0.7045 &0.4611 &0.8756 &0.0151 &0.7723 &0.0076  \\\hline
\textbf{Transfer Models}& & & & &  \\
ESMM&0.7042 &0.4604 &0.8799 &0.0151 &0.7710 &0.0074 \\
AITM&0.7035 &0.4622 &0.8729 &0.0152 &0.7777 &0.0074  \\\hline

\textbf{MTL Backbones}& & & & & &  \\
Shared-Bottom&0.7039 &0.4588 &0.8718 &0.0151 &0.7658 &0.0076 \\
MMOE&0.7051 &0.4570 &0.8780 &0.0150 &0.7792 &0.0074 \\
PLE &0.7041 &0.4587 &0.8787 &0.0150 &0.7769 &0.0075 \\
STEM&0.7046 &0.4584 &0.8804 &0.0150 &0.7803 &0.0075 \\
\hline

\textbf{Ours}& & & & & &\\
Shared-Bottom+EKTM&0.7049 &\textbf{0.4521} &0.8800 &0.0150 &0.7813 &0.0075   \\
MMOE+EKTM &0.7054 &0.4539 &0.8817 &0.0150 &0.7872 &\textbf{0.0073} \\
PLE+EKTM &\textbf{0.7057} &0.4554 &\textbf{0.8835} &0.0150 &\textbf{0.7910} &0.0074 \\
STEM+EKTM &0.7054 &0.4534 &0.8815 &\textbf{0.0149} &0.7893 &0.0075  \\

\bottomrule
\end{tabular}
\end{table*}

\subsubsection{Implementation Details and Evaluation Metrics.}
For the public datasets, we trained four popular multi-task recommendation backbones, namely Shared-Bottom, MMOE and PLE, and STEM. The MMOE model consists of three shared, single-layer fully connected networks as expert network. The PLE model and STEM model include three shared expert networks and one task-specific expert network for each task, with four progressive layers for PLE model. The learning rate was set to 1e-3, weight decay to 1e-6, and the batch size was set to 4096. The loss weights $\alpha$ was grid-searched in the range [0.01, 0.1, 1.0], and the number of heads was grid-searched in the range [1, 2, 4, 8]. The multiple perceptron network for tower architecture of all the MTL methods is set to [512, 256, 128]. To deploy experiments, we utilize a three-node cluster, where each node is with 48 core Intel Xeon CPU E5-2670 (2.30GHZ), 400GB RAM, Linux system, and as well as 2 V100 cards. Besides, 10 times experiments for all methods on each of dataset are conducted. We adopt AUC and LogLoss as the evaluation metric, which are the most widely-used in recommendation field. 
\begin{equation}
    AUC = \frac{\sum_{i=1}^{N_+} \sum_{j=1}^{N_-} \mathbb{I}(p_i > p_j)}{N_+ \times N_-}
\end{equation}
where $N_+$ and $N_-$ denote the number of positive and negative samples, 
$p_i /p_j$ are predicted scores, and $\mathbb{I}$ is an indicator function.
AUC measures the probability of a positive sample being ranked higher than a randomly chosen negative sample. LogLoss, or Binary Cross-Entropy Loss is defined in Equation~\ref{eq:ctr} measures the divergence between predicted probabilities and actual binary labels, serving as a direct indicator of model calibration.

\subsection{Experiments in public datasets}
We first evaluate EKTM on a two-task AliExpress dataset, which consists of CTR and CTCVR prediction tasks.Next, experiments on three-task KuaiVideo dataset are conducted, which contains click, like and follow tasks. The results are shown in Table~\ref{tab:two_tasks} and Table~\ref{tab:three_tasks}  The comparative methods can be categorized into four types. The first type is single-task learning, which employs two separate models for the two tasks. The second type includes typical methods that transfer knowledge both implicitly and explicitly. The third type consists of state-of-the-art multi-task learning backbone methods, which are widely adopted in industrial platforms. The fourth type features backbones integrated with our EKTM method. For AUC metric, the bigger values indicate superior performance; while For LogLoss metric, the smaller values indicate superior performance. We could draw several conclusions as following: EKTM significantly enhances all MTL backbones, proving its adaptability to diverse scenarios. The improvements are consistent across all subsets in AliExpress dataset and KuaiVideo dataset, highlighting its ability to bring beneficial knowledge and mitigate negative transfer. Traditional transfer methods and MTL backbones fail to outperform single task learning model in some scenarios (eg. AliExpress\_US and AliExpress\_ES) , revealing their ineffectiveness in handling task conflicts. These findings underscore the superior design and effectiveness of EKTM in overcoming such challenge.

\subsection{Ablation Study}
The primary goal of the ablation study is to thoroughly assess the individual impact of the custom transmitter module and the calibration loss within the EKTM framework. To optimize resource usage and minimize computational costs, we carried out the experiments using the MMOE model on two specific datasets: AliExpress\_NL and AliExpress\_US. For these experiments, we employed the best configuration settings for MMOE tailored to each dataset, where the parameter 
$\alpha$was set to 0.5, and the number of attention heads was set to 4 for AliExpress\_NL and 8 for AliExpress\_US, respectively. The baseline, labeled "linear," corresponds to a model that exclusively utilizes a simple linear projection mechanism for knowledge transfer without any specialized adaptation. The experimental results, presented in Table~\ref{tab:abl_tasks}, reveal several key observations. Notably, the transmitter module facilitates more effective information transfer compared to the linear projection, thereby helping to alleviate the negative transfer problem to a certain extent. In particular, MMOE augmented with the transmitter module consistently outperforms the linear baseline, with a more pronounced improvement observed on the AliExpress\_NL dataset. Furthermore, the inclusion of the auxiliary calibration loss demonstrates a crucial role in steering the model towards positive optimization for both the transmitter module and the linear layer. This auxiliary loss significantly enhances the performance of MMOE combined with either the linear or transmitter modules. Collectively, these findings underscore the effectiveness and complementary nature of the two core components of EKTM. In summary, the ablation study confirms that EKTM is thoughtfully designed to effectively mitigate the challenges associated with negative transfer, thereby improving overall multi-task learning outcomes.

\subsection{Experiment in industrial dataset}

\begin{table*}[!t]
\caption{Offline experiments on eight-tasks industrial dataset. The AUC metric is adopted as evaluation criterion and the best results are showed in bold.}
\centering
\resizebox{\textwidth}{!}{
\begin{tabular}{c|cccccccc}
\toprule
method&CTR Task&CVR Task 1&CVR Task 2&CVR Task 3&CVR Task 4&CVR Task 5&CVR Task 6&CVR Task 7\\\hline
Baseline&0.91808&0.89130&0.91590&0.93085&0.92787&0.89230&0.90817&0.74356\\
EKTM&\textbf{0.91855}&\textbf{0.89143}&\textbf{0.92010}&\textbf{0.93310}&\textbf{0.92831}&\textbf{0.89290}&\textbf{0.90855}&\textbf{0.74428}\\
\bottomrule
\end{tabular}
}
\label{tab:offline}
\end{table*}

\subsubsection{Offline Experiment}
Offline experiments were conducted using an industrial dataset collected from a large-scale commercial platform, which includes one CTR task and seven CVR tasks, such as activation and payment. To closely reflect real-world recommendation scenarios, the dataset was split chronologically into training (80\%), validation (10\%), and testing (10\%) sets. Model performance was evaluated based on the AUC, maintaining consistency with the evaluation metrics used in experiments on public datasets. The baseline model used for comparison is a highly optimized multi-task learning framework that integrates several widely adopted modules derived from state-of-the-art multi-task learning approaches found in the literature. For training, we set the embedding size to 30, the batch size to 12,000, and the learning rate to 1e-3. The model was trained using the Adam optimizer over 50 epochs, with early stopping triggered by validation loss to prevent overfitting. Table~\ref{tab:offline} provides a detailed performance comparison between the proposed EKTM method and the baseline on the industrial dataset. Our method consistently outperforms the baseline across all tasks, clearly demonstrating its effectiveness and robustness in handling large-scale, real-world data.

\subsubsection{Online Experiment}
\begin{figure}[h]
    \centering
     \includegraphics[scale=0.4]{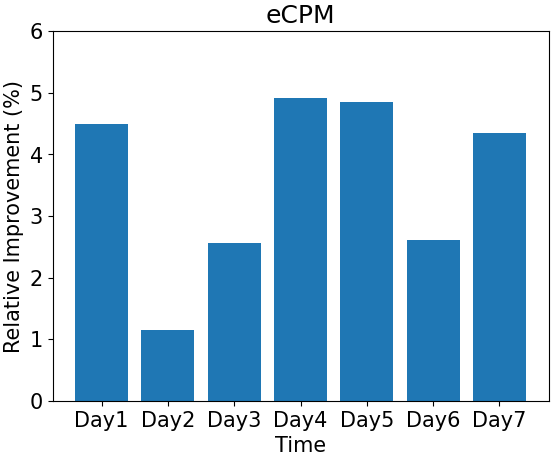}
    \caption{Online A/B testing for our EKTM model versus the baseline model was carried out over a one-week period. The North Star metric eCPM is presented.
    }
    \label{fig:online}
\end{figure}

We further deployed EKTM on a real-world commercial platform that serves hundreds of millions of active users each month to rigorously evaluate its effectiveness within a practical, large-scale environment. 
The baseline model employed for comparison is a highly-optimized, multi-task ranking model that has been meticulously fine-tuned and enhanced over several years, making it a strong standard for benchmarking.
For a controlled online A/B testing experiment conducted on a major channel of the platform, 5\% of the incoming traffic was randomly selected and assigned to the control group, which continued receiving recommendation results generated by the baseline model. Simultaneously, another 5\% of the traffic was rerouted to an experimental group that received results generated by our proposed EKTM method. 
Over the course of the testing period, as continual improvements and validations were made, the proportion of traffic served by the EKTM algorithm was systematically increased in stages: first to 20\%, then to 50\%, and ultimately to 100\% of the traffic. This gradual traffic ramp-up allowed careful monitoring of performance metrics and user engagement to confirm the model's robustness and effectiveness at each stage before full deployment.
The results of this deployment, as illustrated in Figure \ref{fig:online}, demonstrate that the integration of the EKTM algorithm significantly outperforms the baseline method. Specifically, the key North Star metric, eCPM, showed an average improvement of 3.93\%, confirming that EKTM drives better monetization by delivering higher quality recommendations that engage users more effectively. This practical success on a commercial platform highlights EKTM’s strong potential to enhance large-scale content ranking and recommendation systems in dynamic, user-intensive environments.


\section{Conclusion}
In this paper, we propose the Effective Knowledge Transfer Method, a novel approach that enhances multi-task learning in recommendation systems by efficiently transferring relevant knowledge across related tasks. EKTM addresses the challenge of sparse training data, particularly in multiple conversion rate prediction tasks. The core innovation of our method lies in three key components: a centralized router module that seamlessly integrates information across tasks, a transmitter module designed to select task-specific knowledge tailored to each task’s requirements, and an enhanced CVR prediction module that ensures the transferred knowledge contributes positively. Through comprehensive experiments on both public datasets and an industrial platform, complemented by online A/B testing, we demonstrate EKTM’s effectiveness in improving CVR prediction performance. Building on these foundational components, EKTM effectively captures the complex relationships among tasks and dynamically adapts to varying task demands, thereby significantly enhancing the overall robustness and accuracy of multi-task recommendation models in diverse real-world scenarios.

\bibliographystyle{ACM-Reference-Format}
\bibliography{sample-base}


\end{document}